\documentclass[graybox]{svmult}
\usepackage{amsmath}
\usepackage{newtxmath}
\usepackage[latin9]{inputenc}
\usepackage{color}
\usepackage{array}
\usepackage{float}
\usepackage{url}
\usepackage{graphicx}

\makeatletter

\providecommand{\tabularnewline}{\\}

\usepackage{type1cm}
\usepackage{makeidx}
\usepackage{multicol}
\usepackage[bottom]{footmisc}
\usepackage{hyperref}

\usepackage{newtxtext}%
\usepackage[T1]{fontenc}

\makeatother

\begin{document}

\title{Lifting Lockdown Control Measure Assessment: From Finite to Infinite-dimensional
Epidemic Models for COVID-19*}

 \titlerunning{Lifting Lockdown Control Measure Assessment for COVID-19.}

\author{Redouane Qesmi and Aayah Hammoumi}
\institute{Redouane Qesmi \at Superior School of Technology, Sidi Mohamed Ben
Abdellah University, Fez 30000, Morocco, \email{redouane.qesmi@usmba.ac.ma}
\and Aayah Hammoumi \at Department of Biology, Cadi Ayyad University,
Semlalia, Marrakech 40000, Morocco \email{a.hammoumi@uca.ma} \and (*) This is a preprint of a paper whose final and definite form is published in Analysis of Infectious Disease Problems (Covid-19) and Their Global Impact, Springer Nature Singapore Pte Ltd.
Submitted August 14, 2020; revised September 18, 2020; accepted October 8, 2020.}

\maketitle

\chapsubtitle{Lifting Lockdown Control Measure Assessment for COVID-19.}

\abstract*{The main focus of this chapter is on public health control strategies
which are currently the main way to mitigate COVID-19 pandemic. We
introduce and compare compartmental models of increasing complexity
for COVID-19 transmission to describe dynamics of the disease spread.
We begin by considering an SEAIR model including basic characteristics
related to COVID-19. Next, we shall pay attention to age-structure
modeling to emphasis the role of age-group individuals on the disease
spread. A Model with constant delay is also formulated to show the
impact of the latency period on the severity of COVID-19. Since there
is evidence that for COVID-19 disease, important relationships exist
between what is happening in the host and what is occurring at the
population level, we shall link the basic model to in-host dynamics
through the so-called threshold-type delay models. Finally, we will
include demographic effects to the most complex models and we will
conduct rigorous bifurcation analysis to quantify possible factors
responsible for disease progression.}

\abstract{The main focus of this chapter is on public health control strategies
which are currently the main way to mitigate COVID-19 pandemic. We
introduce and compare compartmental models of increasing complexity
for COVID-19 transmission to describe dynamics of the disease spread.
We begin by considering an SEAIR model including basic characteristics
related to COVID-19. Next, we shall pay attention to age-structure
modeling to emphasis the role of age-group individuals on the disease
spread. A Model with constant delay is also formulated to show the
impact of the latency period on the severity of COVID-19. Since there
is evidence that for COVID-19 disease, important relationships exist
between what is happening in the host and what is occurring at the
population level, we shall link the basic model to in-host dynamics
through the so-called threshold-type delay models. Finally, we will
include demographic effects to the most complex models and we will
conduct rigorous bifurcation analysis to quantify possible factors
responsible for disease progression.}

\section{Introduction \label{sec:Introduction}}

The SARS-CoV-2, designated as Severe Acute Respiratory Syndrome CoronaVirus-2,
is a causative agent of COVID-19 disease that first emerged in China
on December 2019 \cite{Ye}. Up to date, $213$ countries and territories
are affected by the disease, with nearly 19 million confirmed cases
and more than 700,000 deaths. Unfortunately, there are no current
effective therapeutic agents or vaccines for treatment of COVID-19
and, consequently, public health control strategies that diminish
contact between infectious and susceptible individuals are actually
the main way to contain and mitigate the pandemic \cite{Wilder}.
Examples of such a control include the personal protective measures
(e.g. hand hygiene, cough or sneeze etiquette and face coverings),
social distancing measures (e.g increasing physical distance from
other people, avoiding social gathering and stay at home) and environmental
surface cleaning \cite{Lasry}. However, despite health control interventions
seem to be successful in reducing the spread of the pandemic, they
are also responsible of global economic crisis. It is estimated that
COVID-19 could cost the global economy more than \$10 trillion \cite{Ahmed,Nicola}.
Millions of individuals throughout the world have been forced to reduce
their working hours or have lost their jobs and around 900 million
students are affected by national school closures \cite{UNESCO1,Viner}.
Furthermore, because of the isolation and the increase of social class
inequalities, the lockdown is badly lived by the majority of children
and adults who have developed mental health disorders and familial
problems \cite{Ahmed,Jakovljevic,Nicola}. Although many information
related to COVID-19, such as its potential to reoccur, remain unclear,
lifting lockdown measure became an urgent need to avoid the worsening
of the global crisis caused by COVID-19 \cite{Prem}. Governments
around the world encountered serious difficulties to adopt the best
lockdown lifting strategy balancing between economy recovery and health
protection of citizens.

Fortunately, since the beginning of the outbreak of COVID-19, mathematical
modeling proved to be an effective tool to predict the course as well
as the severity of the epidemic and to help decision makers to evaluate
the effectiveness of health control measures (See \cite{Hammoumi,Leung,Liu,Ndariou,Vespignani}
and references therein). Generally, at the beginning of an emergent
epidemic, the novel pathogenic agent lacks detailed knowledge. Mathematicians
begin by using simplest compartmental models to estimate the key epidemiological
parameters (such as the basic reproduction number $R_{0}$, peak time,
peak size, latency period, infectious period, etc) which are necessary
to set up public health strategies and monitor the disease progression
\cite{Britton}. Most such models consider individuals in a closed
population that are classified according to their disease status:
susceptible (S), latent or Exposed (E), infectious (I) and recovered
(R). These basic models have the great advantage of being easy to
investigate but, on the other hand, often oversimplify the existing
complexity of disease processes which underestimate or overestimate
the magnitude of the disease spread. However, more complex models
can be considered in case the scientists provide more epidemiological
evidences leading to detailed knowledge of Sars-Cov-2 pathogenic mechanism
and its mode of transmission. Even if they are more difficult to analyze
and need more detailed data, these models could be more realistic
since they take into account more realistic epidemiological properties
such as the heterogeneity of disease transmission, age-subgroups,
latent periods and so on. The purpose of this chapter is to consider
and compare different deterministic compartmental models of increasing
complexity that will be useful to clarify how Sars-CoV-2 spread within
individuals while considering the relaxation of the compulsory lockdown
to prevent dissemination of the COVID-19 disease.

The main contribution of this chapter is organized as follows. In
Section \ref{sec:A-Basic-COVID-19}, we will consider and investigate
an extended basic SEAIR model which is widely used for COVID-19 disease.
Indeed, this model takes into account the standard epidemiological
states such as the exposed individuals to the Sars-CoV-2 virus which
are infected but can not transmit the virus to others. In Section
\ref{sec:Discrete-Age-Structure}, we consider an extension of the
previous model with particular focus on an identified route of COVID-19
transmission from children to adults and vice versa. In other words,
we present a discrete age-structured model, by separating the population
into two different age-subgroups with different contact rates, to
look at the heterogeneity of COVID-19 transmission within a population.
Indeed, it is proved that children are less affected by Sars-Cov-2
than adults and play a minor role in disease transmission \cite{Benjamin,Macartney}.
Seniors and person with existing chronic medical conditions develop
more severe form of disease and are more likely to die \cite{WHO1}.
Furthermore, individuals with different ages may also have different
behaviors and behavioral changes which are crucial in the assessment
of control scenarios targeted at particular groups, such as reopening
schools or relaunch of economic activities. The impact of the latent
period on disease transmission is also evaluated in Section \ref{sec:COVID-19 with delay}
by incorporating a time delay to the basic SAIR model instead of considering
the latent stage as a model component. In Section \ref{sec:COVID-threshold--Model},
we examine a threshold-type delay model by incorporating a series
of smaller Sars-CoV-2 viral loads, due to close contact with infectious
individuals, into the within-host virus dynamics. This type of model
can be used for example to study the impact of COVID-19 exposure to
health care workers who are daily in close contacts with COVID-19
patients, visitors or co-workers in a population and are more likely
to be infected \cite{Heneghan}. In Section \ref{sec:Models-with-Demographic},
we include vital dynamics to the constant and threshold-type delay
models by assuming that the COVID-19 disease could persist for a long
period. Indeed, with the evidence of sensibility of Sars-CoV-2 to
climatic factors such as temperature and humidity, scientists think
that the COVID-19 allows a seasonal cycle and could reduce with climate
change. But, even with the arrival of the warm weather, the disease
still evolved and could persist for several months or years. This
demonstrates that the climate change is not the main parameter which
influences the disease transmission but also, the people\textquoteright s
behavior, the low immunity of individuals to a novel coronavirus and
the immunity period (the amount of time that people remain immune
after infection) \cite{Kanzawa,Kissler}. We will then conduct rigorous
qualitative analysis including bifurcation investigation of both models
with demographic effect to quantify possible factors responsible for
disease progression and highlight long term qualitative behavior of
COVID-19 spread. Next, in order to help scientists to avoid major
blunders and generate models that fit the data reasonably accurately,
we compare in Section \ref{sec:Lifting-Control} the proposed models
to identify which one best fits the reported data and provide a better
prediction for COVID-19. Results of our models will help to tackle
health concerns that are of great importance and will draw of the
hospitals research and surveillance data to create, optimize, and
parameterize disease models, focusing on COVID-19.

\section{Data Collection}

The data of reported symptomatic infectious cases is collected each
day at $11$ pm from the official Coronavirus Portal of Morocco \cite{Ministry9M}.
Data information covers the cumulative number of reported cases from
March $2nd$ to June $10th,$ $2020$. The data from March $2nd$
to March $20$ (first day of lockdown measure) are used to estimate
the basic reproduction number and adjust the investigated models to
become closer to reality, while data from March $21st$ to June $10th$
(last day of lockdown measure) are used to adjust the models and estimate
the lockdown rate during this period of lockdown.

\section{Basic COVID-19 Model \label{sec:A-Basic-COVID-19}}

The population, with size $N,$ considered in this basic model is
stratified into seven disease status. Individuals are classified as
susceptible $(S)$, exposed noninfectious $\left(E\right),$ asymptomatic
infectious $(A)$, unreported symptomatic infectious $(I_{u}),$ reported
symptomatic infectious or hospitalized $(H),$ recovered $(R)$ and
dead $(D).$ We formulate the model to describe the course of COVID-19
epidemic under the assumptions:
\begin{enumerate}
\item Reported symptomatic infectious individuals are hospitalized and can
not contact susceptibles anymore.
\item As confirmed by Rothe et al. \cite{Rothe}, asymptomatic  individuals
can infect susceptible individuals.
\item Confined asymptomatic and confined unreported individuals can still
spread the virus to their families.
\item Exposed infected individuals can not immediately spread the virus
to other individuals.
\item As proved by MacIntyre in \cite{MacIntyre}, asymptomatic and symptomatic
infectious individuals share the same infection probability.
\end{enumerate}
Taking account of the previous assumptions, the dynamics of COVID-19
can be described as follows: Individuals are confined at rate $p.$
Unconfined (resp. confined) susceptibles $(1-p)S$ (resp. $(pS)$
) contacted with either unreported symptomatic $(I_{u})$ or asymptomatic
infectious individuals $(A)$ are infected with infection probability,
$\beta_{N}$ (resp. $\beta_{c}$), and move to the exposed infected
class $(E).$ Exposed individuals then become asymptomatic infectious
at rate $k.$ After an average period $1/\delta$ days the asymptomatic
infectious individuals $(A)$ become symptomatic and proceed either
to the unreported symptomatic infectious $(I_{u}),$ at rate $\delta_{1},$
or to the reported symptomatic infectious $(H)$ at rate $\delta_{2}$
with $\delta=\delta_{1}+\delta_{2}.$ Once becoming symptomatic, individuals
of class $I_{u}$ and $H$ remain symptomatic for $1/\mu$ days on
average before they are recovered or dead at rate $d.$ The parameter
$\gamma$ corresponds to the lifting rate while parameter $\theta$
corresponds to the contact reduction, due to wearing masks, washing
hands, and social distancing practices of unconfined individuals.
The general basic model equations including parameters control is
given as follow
\begin{equation}
\begin{cases}
{\displaystyle \frac{dS}{dt}} & =-\left(\left(1-\gamma\right)p\beta_{c}+(1-\theta)\left(1-\left(1-\gamma\right)p\right)\beta_{N}\right)S(t)\left(A(t)+I_{u}(t)\right)/N,\vspace{0.05in}\\
{\displaystyle \frac{dE}{dt}} & =\left(\left(1-\gamma\right)p\beta_{c}+(1-\theta)\left(1-\left(1-\gamma\right)p\right)\beta_{N}\right)S(t)\left(A(t)+I_{u}(t)\right)/N-kE,\\
{\displaystyle \frac{dA}{dt}} & =kE-\delta A(t),\vspace{0.05in}\\
{\displaystyle \frac{dI_{u}}{dt}} & =\delta_{1}A(t)-\mu I_{u}(t)-dI_{u}(t),\vspace{0.05in}\\
\frac{dH}{dt} & =\delta_{2}A(t)-\mu H-dH,\vspace{0.05in}\\
{\displaystyle \frac{dR}{dt}} & =\mu(H+I_{u}),\vspace{0.05in}\\
{\displaystyle \frac{dD}{dt}} & =d(H+I_{u}).\vspace{0.05in}
\end{cases}\label{eq:ODE}
\end{equation}

\subsection{Reproduction Numbers}

The basic reproduction number, $R_{0},$ is the average number of
secondary infections produced when one infectious individual is introduced
into a host susceptible population. This quantity determines whether
a given disease may spread, or die out in a population. To compute
this number, we assume that $p=\theta=\gamma=0$ and we apply the
next generation matrix method in \cite{VandenDriessche}. We obtain
\begin{equation}
R_{0}\mbox{=}\frac{\left(a+k\right)\left({\displaystyle {\displaystyle a}}+\delta\right)\left({\displaystyle a}+\mu+d\right)}{k\left(\delta_{1}+{\displaystyle a}+\mu+d\right)}\left(\frac{1}{\delta}+\frac{\delta_{1}}{\delta\mu}\right)\label{eq:R0-Basic}
\end{equation}
where $a$ is an estimated constant given in Subsection \ref{subsec:Parameter-and-initial}.
Here, $R_{0}$ can be explained as follows: Assume that one asymptomatic
infectious individual is introduced into the susceptible population.
This asymptomatic individual produces, on average, $\beta_{N}S_{0}{\displaystyle \frac{1}{\delta}}$
asymptomatic individuals during his average lifespan $1/\delta.$
These asymptomatic individuals then become unreported symptomatic
infectious individuals over their lifespan $1/\delta$ at a rate $\delta_{1}$
and then each infectious symptomatic produces, on average, $\beta_{N}S_{0}{\displaystyle \frac{1}{\mu}}$
asymptomatic individuals during his lifespan $1/\mu$.

Let us show the formula of $R_{0}.$ The linearized system related
to infectious individuals , around $\left(S_{0},0,0,0\right),$ of
system (\ref{eq:ODE}) is given by
\[
\begin{cases}
\frac{dE}{dt} & =-kE(t)+\beta_{N}S_{0}A(t)+\beta_{N}S_{0}I_{u}(t)\\
{\displaystyle \frac{dA}{dt}} & =kE(t)-\delta A(t),\vspace{0.05in}\\
{\displaystyle \frac{dI_{u}}{dt}} & =\delta_{1}A(t)-\left(\mu+d\right)I_{u}(t),\vspace{0.05in}
\end{cases}
\]
and the associated Jacobian matrix is given by $M=F-E$ where
\[
F=\left(\begin{array}{ccc}
0 & \beta_{N}S_{0} & \beta_{N}S_{0}\\
k & 0 & 0\\
0 & \delta_{1} & 0
\end{array}\right)\mbox{ and }E=\left(\begin{array}{ccc}
k & 0 & 0\\
0 & \delta & 0\\
0 & 0 & \mu+d
\end{array}\right).
\]
Therefore, $FE^{-1}=\left(\begin{array}{ccc}
0 & \beta_{N}S_{0}/\delta & \beta_{N}S_{0}/\left(\mu+d\right)\\
1 & 0 & 0\\
0 & \delta_{1}/\delta & 0
\end{array}\right)$ and $R_{0}$ is its spectral radius. Using the formula of $\beta_{N}$
in (\ref{eq:5-1-1}), we obtain the formula given in (\ref{eq:R0-Basic}).

The control reproduction number, $R_{c},$ is an important value,
used to determine whether a control policy, such as lockdown, lifting,
behavioral practices, etc, will be efficient to decrease the number
of secondary infections to be less than one. Computation method of
$R_{c}$ is similar to the one of $R_{0}$ and leads to the following
formula
\[
R_{c}=\left(\left(1-\gamma\right)p\beta_{c}+(1-\theta)\left(1-\left(1-\gamma\right)p\right)\beta_{N}\right)\left(\frac{1}{\delta}+\frac{\delta_{1}}{\delta\mu}\right).
\]

\subsection{Parameter and Initial Data Estimation \label{subsec:Parameter-and-initial}}

To estimate the model parameters we will consider two different stages.
The first stage is between the beginning of the COVID-19 epidemic
and the first time of containment control (i.e $p=0$ and $\theta=0$)
for which we will estimate the initial data of the model, the parameters
related to infection and the basic reproduction number. The second
stage will be during the lockdown period ($\gamma=0$ and $\theta=0$)
for which we will estimate the lockdown rate.

Since the first and the only symptomatic infectious individual is
reported on March $2nd$, $2020$, which corresponds to $t=0,$ then
$H(0)=1,$ $R(0)=0$ and $D(0)=0.$ For the estimation of $\beta_{N},$
$E(0),$ $A(0)$ and $I_{u}(0)$ we will use the data of cumulative
reported cases collected from March 2nd to March 20 (before the start
of lockdown) and we follow the procedure by \cite{Liu}. The cumulative
reported infectious population is given, for $t\geq0,$ by $F(t)=\delta_{2}\int_{0}^{t}A(s)ds+1$.
It is obvious that cumulative reported infectious population increases
slowly and then accelerates rapidly with time. Hence, we will use
exponential regression with $95\%$ of confidence level to find an
exponential function that best fits the data, from March $2nd$ to
June $10th.$ Using SPSS software (Statistical Package for the Social
Sciences) we found that exponential model given by $be^{at}$ with
$\ensuremath{a=0.263}$ with confidence interval $CI\left(0.229-0.297\right)$
and $b=0.507$ with $CI\left(0.3444-0.7475\right)$ fits well the
data with a correlation coefficient given by $R=0.97.$ It follows
from $F(t)=\delta_{2}\int_{0}^{t}A(s)ds+1=be^{at}$ that
\begin{equation}
A(t)={\displaystyle \frac{ba}{\delta_{2}}}e^{at}.\label{eq:tildeA}
\end{equation}
Since the initial susceptible population is not dramatically affected
in the early phase of the epidemic, we will assume that $S(t)\approx S(0).$
Let $S_{0}:=S(0),$ $E(0):=E_{0},$ $A(0):=A_{0}$ and $I_{u}(0):=I_{0}.$
From the second and the third equations of system (\ref{eq:ODE})
and using (\ref{eq:tildeA}) we obtain
\begin{equation}
\left(\left(\frac{{\displaystyle {\displaystyle a}}+\delta}{k}\right)+\left({\displaystyle {\displaystyle a}}+\delta\right)\right)A(t)=\beta_{N}S(0)\left(A(t)+I_{u}(t)\right),\label{eq:estimA-1}
\end{equation}
\begin{equation}
E(t)=E_{0}e^{at}\mbox{ and }I_{u}(t)=I_{0}e^{at},\label{eq:estimI-1}
\end{equation}
where
\begin{equation}
E_{0}=\frac{{\displaystyle {\displaystyle a}}+\delta}{k}{\displaystyle \frac{ba}{\delta_{2}}}\mbox{ and }I_{0}=\frac{ba\left(\left(\frac{{\displaystyle {\displaystyle a}}+\delta}{k}\right)+\left({\displaystyle {\displaystyle a}}+\delta\right)-\beta_{N}S(0)\right)}{\delta_{2}\beta_{N}S_{0}}.\label{eq:4-1}
\end{equation}
Now, using formulas (\ref{eq:estimI-1}) and the third equation of
system (\ref{eq:ODE}), we obtain after simplification
\begin{equation}
{\displaystyle aE_{0}}=\beta_{N}S_{0}\left(A_{0}+I_{0}\right)-kE_{0}\label{eq:5-1}
\end{equation}
and
\begin{equation}
{\displaystyle aI_{0}}=\delta_{1}A_{0}-(\mu+d)I_{0}.\label{eq:6-2}
\end{equation}
Solving equations (\ref{eq:4-1}), (\ref{eq:5-1}) and (\ref{eq:6-2})
for $\beta_{N}$ and $I_{0}$ lead to
\begin{equation}
\beta_{N}={\displaystyle \frac{\left(a+k\right)\left({\displaystyle {\displaystyle a}}+\delta\right)\left({\displaystyle a}+\mu+d\right)}{k\left(\delta_{1}+{\displaystyle a}+\mu+d\right)}}\mbox{ and }I_{0}=\frac{\delta_{1}}{{\displaystyle a}+\mu+d}A_{0}.\label{eq:5-1-1}
\end{equation}
To estimate the transmission rate, $\beta_{c},$ and the lockdown
rate, $p$ during the lockdown period, we assume that $\gamma=0$
and $\theta=0$ and we use the nonlinear least squares solver \textquotedblleft lsqcurvefit\textquotedblright{}
in MATLAB R2019b software. The values of the estimated parameters
are summarized in Table \ref{parametervalues}.

Define the sum of squared residuals ($SSR$) as
\[
SSR=\sqrt{\frac{1}{n}\sum_{i=1}^{n}\left(\delta_{2}A(t)-\mbox{Newcase}(i)\right)^{2}}
\]
where Newcase$\left(i\right)$ is the number of new reported cases
on the day $i$ and $n$ is the number of collected new cases. This
number measures the discrepancy between the data and the estimation
model of new reported cases per day and will serve us to compare the
suggested models of this chapter. A small $SSR$ indicates a better
fit of the model to the data. A computation of this measure for model
(\ref{eq:ODE}) leads to $SSR_{basic}=93.4.$
\begin{table}[!h]
\caption{Parameter definitions and values of model (\ref{eq:ODE}).}
\textsf{\small{}\label{parametervalues}}%
\begin{tabular}{>{\raggedright}p{1.5cm}>{\raggedright}p{4cm}p{2cm}>{\raggedright}p{2cm}>{\raggedright}p{1.7cm}}
\hline
\noalign{\smallskip{}
} Symbol & Definition & Parameter value & Confidence interval & Reference\tabularnewline
\noalign{\smallskip{}
}\svhline\noalign{\smallskip{}
} {\footnotesize{}$S(0)$} & \textsf{I}nitial susceptible popu\textsf{l}ation & $35865191$ &  & \cite{HCP}\tabularnewline
{\footnotesize{}$E(0)$} & Initial exposed noninfecious population & $0.3175$ & $00.8-1.37$ & Estimated\tabularnewline
{\footnotesize{}$A(0)$} & Initial asymptomatic population & $11.9921$ & $9.42-15.03$ & Estimated\tabularnewline
{\footnotesize{}$I_{u}(0)$} & Initial unreported symptomatic population & $0.8414$ & $0.592-1.3$ & Estimated\tabularnewline
{\footnotesize{}$H(0)$} & Initial reported symptomatic population & $1$ &  & See text\tabularnewline
{\footnotesize{}$R(0)$} & Initial recovered population & $0$ &  & See text\tabularnewline
{\footnotesize{}$D(0)$} & Initial dead population & $0$ &  & See text\tabularnewline
\textsf{\small{}$\beta_{N}$} & Infection rate for unconfined population & $2.87$ & $0.5-4.2$ & Estimated\tabularnewline
\textsf{\small{}$\beta_{c}$} & Infection rate for confined population & $0.57$ & $0.1-0.84$ & Estimated\tabularnewline
\textsf{\small{}$1/\delta$} & Asymptomatic duration & $6$ days &  & \cite{Ministry9M}\tabularnewline
$k$ & Exposed noninfectious rate & $3$ & $2-4$ & See text\tabularnewline
\textsf{\small{}$\delta_{1}$} & Asymptomatic unreported rate & $0.017$ per day &  & Assumed\tabularnewline
\textsf{\small{}$\delta_{2}$} & Symptomatic reported rate & $0.15$ per day &  & Assumed\tabularnewline
\textsf{\small{}$1/\mu$} & Symptomatic duration & $14$ days &  & \cite{WHO25}\tabularnewline
{\small{}$p$} & Proportion of lockdown & $0.7$ & $0.5-0.76$ & Estimated\tabularnewline
{\small{}$R_{0}$} & Basic reproduction number & $2.88$ & $2.55-2.99$ & Estimated\tabularnewline
\noalign{\smallskip{}
}\noalign{\smallskip{}
} &  &  &  & \tabularnewline
\end{tabular}
\end{table}

\section{Discrete Age Structure COVID-19 Model \label{sec:Discrete-Age-Structure}}

Basic discrete age-structured compartmental models seems to be more
appropriate for COVID-19 disease since it is claimed that adults have
a greater risk of transmitting SARS-CoV-2 virus than children do toward
susceptibles (See Section \ref{sec:Introduction}). This suggest that
in order to give more appropriate description of COVID-19 transmission
it is important to separate the population into two different age-subgroups.

The population considered in this section is stratified into two age
categories and ten disease status. Individuals are classified as susceptible
children $(T)$, susceptible adult $(S)$, exposed noninfectious $\left(E_{T}\right),$
exposed noninfectious adult $\left(E_{s}\right),$ asymptomatic infectious
adult $(A)$, asymptomatic infectious children $(B)$, unreported
symptomatic infectious $(I_{u})$, hospitalized symptomatic infectious
$(H),$ recovered individuals $(R)$ and dead individuals $(D)$.
We assume that infected children do not show symptoms and can still
transmit the disease. COVID-19 disease dynamics can be described as
follows: Let $\beta^{\chi}$ be the transmission rate from infectious
individuals to confined susceptible individuals and $\beta^{N}$ be
the transmission rate from infectious individuals to unconfined susceptible
individuals. Then, for $i\in\left\{ \chi,N\right\} ,$ susceptibles
adults $(S)$ (resp. susceptible children $(T)$) are infected through
contact with infectious adults $(A+I_{u})$ at a transmission rate
$\beta_{aa}^{i}$ (resp. $\beta_{ca}^{i}$) or through contact with
infectious children $(B)$ at a transmission rate $\beta_{ac}^{i}$
(resp. $\beta_{cc}^{i}$) and move to the exposed noninfectious adult
class $(E_{s})$ (resp. the exposed noninfectious children class $(E_{T})$
). Adult exposed individuals (resp. children exposed individuals)
then become asymptomatic infectious at rate $k_{s}$ (resp. $k_{T}$).
After an average period $1/\delta$ days the asymptomatic infectious
individuals $(A)$ become symptomatic and proceed either to the unreported
symptomatic infectious $(I_{u}),$ at rate $\delta_{1},$ or to the
hospitalized individual $(H)$ at rate $\delta_{2}$ with $\delta=\delta_{1}+\delta_{2}.$
Once becoming symptomatic, individuals of class $I_{u}$ and $H$
either remain asymptomatic for $1/\mu$ days on average before they
are recovered or remain asymptomatic for $1/d$ days on average before
they are dead. Asymptomatic children can either be recovered without
being hospitalized at rate $\delta$ or detected and hospitalized
at rate $\sigma.$ The control parameters are as defined in Section
\ref{sec:A-Basic-COVID-19}. The subscripts $c$ and $a,$ respectively,
characterize children and adults. The model will be given by the following
equations
\begin{equation}
\begin{cases}
{\displaystyle \frac{dT}{dt}} & =-\left(1-\gamma_{c}\right)p_{c}\left(\beta_{ac}^{\chi}T(t)\left(A(t)+I_{u}(t)\right)+\beta_{cc}^{\chi}T(t)B(t)\right)\\
 & -\left(1-\theta_{c}\right)\left(1-\left(1-\gamma_{c}\right)p_{c}\right)\left(\beta_{ac}^{N}T(t)\left(A(t)+I_{u}(t)\right)+\beta_{cc}^{N}T(t)B(t)\right),\vspace{0.05in}\\
{\displaystyle \frac{dE_{T}}{dt}} & =\left(1-\gamma_{c}\right)p_{c}\left(\beta_{ac}^{\chi}T(t)\left(A(t)+I_{u}(t)\right)+\beta_{cc}^{\chi}T(t)B(t)\right)\\
 & +\left(1-\theta_{c}\right)\left(1-\left(1-\gamma_{c}\right)p_{c}\right)\left(\beta_{ac}^{N}T(t)\left(A(t)+I_{u}(t)\right)+\beta_{cc}^{N}T(t)B(t)\right)-k_{T}E_{T},\vspace{0.05in}\\
{\displaystyle \frac{dB}{dt}} & =k_{T}E_{T}-(\sigma+\delta)B(t),\vspace{0.05in}\\
{\displaystyle \frac{dS}{dt}} & =-\left(1-\gamma_{a}\right)p_{a}\left(\beta_{aa}^{\chi}S(t)\left(A(t)+I_{u}(t)\right)+\beta_{ca}^{\chi}S(t)B(t)\right)\\
 & -\left(1-\theta_{a}\right)\left(1-\left(1-\gamma_{a}\right)p_{a}\right)\left(\beta_{aa}^{N}S(t)\left(A(t)+I_{u}(t)\right)+\beta_{ca}^{N}S(t)B(t)\right),\vspace{0.05in}\\
{\displaystyle \frac{dE_{s}}{dt}} & =\left(1-\gamma_{a}\right)p_{a}\left(\beta_{aa}^{\chi}S(t)\left(A(t)+I_{u}(t)\right)+\beta_{ca}^{\chi}S(t)B(t)\right)\\
 & +\left(1-\theta_{a}\right)\left(1-\left(1-\gamma_{a}\right)p_{a}\right)\left(\beta_{aa}^{N}S(t)\left(A(t)+I_{u}(t)\right)+\beta_{ca}^{N}S(t)B(t)\right)-k_{s}E_{s},\vspace{0.05in}\\
{\displaystyle \frac{dA}{dt}} & =k_{s}E_{s}-\delta A(t),\vspace{0.05in}\\
{\displaystyle \frac{dI_{u}}{dt}} & =\delta_{1}A(t)-\mu I_{u}(t)-dI_{u}(t),\vspace{0.05in}\\
{\displaystyle \frac{dH}{dt}} & =\delta_{2}A(t)+\sigma B(t)-\mu H-dH,\vspace{0.05in}\\
{\displaystyle \frac{dR}{dt}} & =\delta B(t)+\mu(H+I_{u}),\vspace{0.05in}\\
{\displaystyle \frac{dD}{dt}} & =d(H+I_{u}).
\end{cases}\label{eq:Age}
\end{equation}

\subsection{Reproduction Numbers}

Here, the basic and control reproduction numbers will be given by
\begin{equation}
R_{0}=\frac{\beta_{cc}^{N}T(0)/N}{\sigma+\delta}+\frac{\beta_{aa}^{N}S(0)/N}{\delta}+\frac{\delta_{1}\beta_{aa}^{N}S(0)/N}{\delta\left(\mu+d\right)}\label{eq:R0-age}
\end{equation}
and
\begin{eqnarray}
R_{c} & = & \frac{\left(\left(1-\gamma_{c}\right)p_{c}\beta_{cc}^{\chi}+\left(1-\theta_{c}\right)\left(1-\left(1-\gamma_{c}\right)p_{c}\right)\beta_{cc}^{N}\right)T(0)/N}{\sigma+\delta}\label{eq:Rc-age}\\
 &  & +\frac{\left(\delta_{1}+\mu+d\right)\left(\left(1-\gamma_{a}\right)p_{a}\beta_{aa}^{\chi}+\left(1-\theta_{a}\right)\left(1-\left(1-\gamma_{a}\right)p_{a}\right)\beta_{aa}^{N}\right)S(0)/N}{\delta\left(\mu+d\right)}.\nonumber
\end{eqnarray}
Let us show the formula of $R_{0.}$ By setting $\gamma=\theta=0,$
the linearized system related to infectious individuals, around $\left(T(0),0,0,S_{0},0,0,0\right),$
of system (\ref{sec:A-Basic-COVID-19}) is given b
\[
\begin{cases}
{\displaystyle \frac{dE_{T}}{dt}} & =-k_{T}E_{T}+\beta_{cc}^{N}T(0)B(t)/N+\beta_{ac}^{N}T(0)A(t)/N+\beta_{ac}^{N}T(0)I_{u}(t)/N\vspace{0.05in}\\
{\displaystyle \frac{dB}{dt}} & =k_{T}E_{T}-(\sigma+\delta)B(t),\vspace{0.05in}\\
{\displaystyle \frac{dE_{S}}{dt}} & =\beta_{ca}^{N}S(0)B(t)/N-k_{s}E_{S}+\beta_{aa}^{N}S(0)A(t)/N+\beta_{aa}^{N}S(0)I_{u}(t)/N\vspace{0.05in}\\
{\displaystyle \frac{dA}{dt}} & =k_{s}E_{s}-\delta A(t),\vspace{0.05in}\\
{\displaystyle \frac{dI_{u}}{dt}} & =\delta_{1}A(t)-\left(\mu+d\right)I_{u}(t).
\end{cases}
\]
Moreover, the associated Jacobian matrix will be given by $M=F-E$
where
\[
F=\left(\begin{array}{ccccc}
0 & \beta_{cc}^{N}T(0)/N & 0 & \beta_{ac}^{N}T(0)/N & \beta_{ac}^{N}T(0)/N\\
k_{T} & 0 & 0 & 0 & 0\\
0 & \beta_{ca}^{N}S(0)/N & 0 & \beta_{aa}^{N}S(0)/N & \beta_{aa}^{N}S(0)/N\\
0 & 0 & k_{s} & 0 & 0\\
0 & 0 & 0 & \delta_{1} & 0
\end{array}\right)\mbox{ and }E=\left(\begin{array}{ccccc}
k_{T} & 0 & 0 & 0 & 0\\
0 & \sigma+\delta & 0 & 0 & 0\\
0 & 0 & k_{s} & 0 & 0\\
0 & 0 & 0 & \delta & 0\\
0 & 0 & 0 & 0 & \mu+d
\end{array}\right).
\]
Therefore,
\[
FE^{-1}={\displaystyle \frac{1}{N}{\displaystyle \begin{pmatrix}{\displaystyle \frac{\beta_{cc}^{N}T(0)}{\delta+\sigma}} & {\displaystyle \frac{\alpha_{c}\beta_{cc}T(0)}{\delta+\sigma}} & {\displaystyle \frac{\left(d+\mu+\delta_{1}\right)\beta_{ac}^{N}T(0)}{\delta(d+\mu)}} & {\displaystyle \frac{\left(d+\mu+\delta_{1}\right)\beta_{ac}^{N}T(0)}{\delta(d+\mu)}} & {\displaystyle \frac{\beta_{ac}^{N}T(0)}{d+\mu}}\\
0 & 0 & 0 & 0 & 0\\
{\displaystyle \frac{\beta_{ca}^{N}S(0)}{\delta+\sigma}} & {\displaystyle \frac{\beta_{ca}^{N}S(0)}{\delta+\sigma}} & {\displaystyle \frac{\left(d+\mu+\delta_{1}\right)\beta_{aa}^{N}S(0)}{\delta(d+\mu)}} & {\displaystyle \frac{\left(d+\mu+\delta_{1}\right)\beta_{aa}^{N}S(0)}{\delta\left(d+\mu\right)}} & {\displaystyle \frac{\beta_{aa}^{N}S(0)}{d+\mu}}\\
0 & 0 & 0 & 0 & 0\\
0 & 0 & 0 & 0 & 0
\end{pmatrix}}}
\]
and $R_{0}$ is its spectral radius which is given by formula (\ref{eq:R0-age}).
Moreover, assuming that $\gamma\theta_{c}\theta_{a}\neq0$ and following
the same process above, the control reproduction number will be given
by formula (\ref{eq:Rc-age}).

\subsection{Parameter and Initial Data Estimation}

Note that the first infected child was reported $22$ days since the
beginning of the epidemic. Furthermore, the maximum asymptomatic duration
including the exposure period is about $14$ days. Consequently, there
were neither exposed nor asymptomatic infected children under 15 years
old at $t=0.$ Thus, the initial data values related to infected adult
individuals are the same as those in system \ref{sec:A-Basic-COVID-19}.
Furthermore, $E_{0}=B_{0}=0,$ $T_{0}=9683602$ and $S_{0}=26181589.$
As mentioned in Section \ref{sec:Introduction}, we assume that $\beta_{ac}^{i}=\beta_{aa}^{i},$
$\beta_{ca}^{i}=\beta_{cc}^{i}$ where $i\in\left\{ \chi,N\right\} .$
Using the same fitting solver as in Section \ref{subsec:Parameter-and-initial},
we obtain the parameter values shown in Table \ref{parametervalues-1-1}.
The $SSR$ related to this model is estimated to be $SSR_{age}=98.76.$
\begin{table}[t]
\caption{Parameter definitions and values of model (\ref{eq:Age}).}
\textsf{\small{}\label{parametervalues-1-1}}%
\begin{tabular}{>{\raggedright}p{1.5cm}>{\raggedright}p{5cm}p{2cm}>{\raggedright}p{2cm}>{\raggedright}p{1.6cm}}
\hline
\noalign{\smallskip{}
} Symbol & Definition & Parameter value & Confidence interval & Reference\tabularnewline
\noalign{\smallskip{}
}\svhline\noalign{\smallskip{}
} $\beta_{aa}^{N},\beta_{ac}^{N}$ & Infection rate from infectious adults to unconfined population & $0.41$ & $0.1-0.63$ & Estimated\tabularnewline
$\beta_{ca}^{N},\beta_{cc}^{N}$ & Infection rate from infectious children to unconfined population & $0.2$ & $0.05-0.31$ & Estimated\tabularnewline
$\beta_{aa}^{\chi},\beta_{ac}^{\chi}$ & Infection rate from infectious adults to confined population & $0.1$  & $0.007-0.15$ & Estimated\tabularnewline
$\beta_{ca}^{\chi},\beta_{cc}^{\chi}$ & Infection rate from infectious children to confined population & $0.014$  & $0.002-0.16$ & Estimated\tabularnewline
$p$ & Proportion of lockdown & $0.65$ & $0.57-0.86$ & Estimated\tabularnewline
$k$ & Exposed noninfectious individuals & $3$ & $2-4$ & See text\tabularnewline
$R_{0}$ & Basic reproduction number & $2.06$ & $0.5-3.19$ & Estimated\tabularnewline
\noalign{\smallskip{}
}\noalign{\smallskip{}
} &  &  &  & \tabularnewline
\end{tabular}
\end{table}

\section{COVID-19 Model with Constant Delay \label{sec:COVID-19 with delay}}

In order to enable the study of the effect of the period time, in
which infected individuals are asymptomatic and noninfectious, on
the COVID-19 dynamics, we will incorporate the time delay (latency
period) in the basic model instead of considering the noninfectious
latent state as a model component. Let $\eta$ denotes the death rate
of noninfectious exposed individuals. Once infected through contact
with infectious individuals at rate $\beta,$ the susceptible individuals
that survive with probability $e^{-\eta\tau}$ become infectious (able
to transmit the infection) when the time since exposure exceeds an
exposure period time $\tau.$ The dynamics of the model are described
by the following system of differential equation with delay

\begin{equation}
\begin{cases}
\frac{dS}{dt} & =-\alpha e^{-\eta\tau}S(t)\left(A(t)+I_{u}(t)\right)/N,\vspace{0.05in}\\
{\displaystyle \frac{dA}{dt}} & =\alpha e^{-\eta_{a}\tau}S(t-\tau)\left(A(t-\tau)+I_{u}(t-\tau)\right)-\delta A(t),\vspace{0.05in}\\
{\displaystyle \frac{dI_{u}}{dt}} & =\delta_{1}A(t)-\mu I_{u}(t)-dI_{u}(t),\vspace{0.05in}\\
\frac{dH}{dt} & =\delta_{2}A(t)-\mu H-dH,\vspace{0.05in}\\
{\displaystyle \frac{dR}{dt}} & =\mu(H+I_{u}),\vspace{0.05in}\\
{\displaystyle \frac{dD}{dt}} & =d(H+I_{u})
\end{cases}\label{eq:delay}
\end{equation}
where $\alpha=\left(1-\gamma\right)p\beta_{c}+(1-\theta)\left(1-\left(1-\gamma\right)p\right)\beta_{N}$
and all the model parameters, except $\eta$ and $\tau,$ are described
similarly to those in Section \ref{sec:A-Basic-COVID-19}.

\subsection{Reproduction Numbers \label{subsec:Basic-reproduction-number delay}}

The control and basic reproduction numbers for system (\ref{eq:delay})
are, successively given by

\[
R_{c}\mbox{=}\frac{\alpha S_{0}e^{-\eta\tau}}{N}\left(\frac{1}{\delta}+\frac{\delta_{1}}{\delta\left(\mu+d\right)}\right)\mbox{ and }R_{0}\mbox{=}\frac{\beta_{N}S_{0}e^{-\eta\tau}}{N}\left(\frac{1}{\delta}+\frac{\delta_{1}}{\delta\left(\mu+d\right)}\right).
\]
To compute the Basic reproduction number, we apply the survival function
approach described by Heffernan, Smith, and Wahl \cite{Heffernen2005}.
Let $R_{01}$ (respectively, $R_{02}$) be the average number of secondary
infections produced when one asymptomatic infected (respectively,
symptomatic unreported infected) individual is introduced into the
host virgin population. Following the work in \cite{Heffernen2005},
we have $R_{01}=\int_{0}^{\infty}F(s)ds$ where $F(s)$ is the probability
that a newly asymptomatic infected individual has been produced by
an existing asymptomatic infectious individual and lives for at least
time $s$. The probability function $F(s)$ can be expressed as $F(s)={\displaystyle \int_{0}^{s}P_{1}(t)P_{2}(s,t)dt,}$
where $P_{1}(t)$ is the probability that an asymptomatic infected
individual of age $t$ infects a susceptible individual and is given
by ${\displaystyle \frac{\beta_{N}S_{0}}{N}}$ and $P_{2}\left(s,t\right)$
is the probability that exposed infected individual lives to age $s-t$
and is given by $e^{-\eta\tau}e^{-\delta(s-t)}$ before becoming infectious.
Consequently $R_{01}=\frac{\beta_{N}S_{0}}{N}e^{-\eta\tau}\int_{0}^{\infty}\int_{0}^{s}e^{-\delta(s-t)}dtds$
which can be reduced to $R_{01}=\frac{\beta_{N}S_{0}e^{-\eta\tau}}{N\delta}.$Since
asymptomatic individuals can become unreported symptomatic infectious
individuals over their lifespan $1/\delta$ at a rate $\delta_{1}$
and each infectious symptomatic individual produces, on average, $\beta_{N}S_{0}{\displaystyle \frac{1}{\mu+d}}$
asymptomatic individuals during his lifespan $1/\left(\mu+d\right)$
then, similarly, we can express $R_{02}$ as $R_{02}\mbox{=}\frac{\beta_{N}S_{0}e^{-\eta\tau}}{N}\frac{\delta_{1}}{\delta\left(\mu+d\right)}.$
Thus, the basic reproduction number $R_{0}$ will be given by
\[
R_{0}\mbox{=}\frac{\beta_{N}S_{0}e^{-\eta\tau}}{N}\left(\frac{1}{\delta}+\frac{\delta_{1}}{\delta\left(\mu+d\right)}\right).
\]
Similarly, we can obtain the control reproduction number for system
(\ref{eq:delay}).

\subsection{Parameter and Initial Data Estimation \label{subsec:Parameters-and-initial}}

Since the parameters $\delta,$ $\delta_{1},$$\delta_{2},$ $\mu$
and $d$ are not affected by the age then their values are the same
as those in Subsection \ref{subsec:Parameter-and-initial}. Note that
since there were no death of exposed individuals then it is meaningful
to assume that $\eta=0.$ To estimate the initial data we will use
the same process as in Subsection \ref{subsec:Parameter-and-initial}
so that $A(t)$ will be given by (\ref{eq:tildeA}) for $t$ close
to $0.$ Thus, using the second and the third equations of system
(\ref{eq:delay}) we obtain, for $t$ close to $0,$

\begin{equation}
{\displaystyle aA(t)}=\beta_{N}S(0)\left(A(t-\tau)+I_{u}(t-\tau)\right)/N-\delta A(t),\label{eq:estimA-1-1}
\end{equation}
 and

\begin{equation}
I_{u}(t+\theta)=I_{0}\left(\theta\right)e^{at},\label{eq:estimI-1-1}
\end{equation}
 where

\begin{equation}
{\displaystyle I_{0}(\theta)=\frac{\delta_{1}}{{\displaystyle a}+\mu+d}{\displaystyle \frac{ba}{\delta_{2}}}e^{a\theta}}.\label{eq:4-1-1}
\end{equation}
Now, using equations (\ref{eq:estimA-1-1}) and (\ref{eq:estimI-1-1})
and the third equation of system (\ref{eq:delay}), we obtain after
simplification

\begin{equation}
{\displaystyle aA_{0}}=\beta_{N}S_{0}\left(A_{0}+\frac{\delta_{1}}{a+\mu+d}A_{0}\right)e^{-a\tau}/N-\delta A_{0}\label{eq:5-1-2}
\end{equation}
 and
\begin{equation}
{\displaystyle aI_{0}}=\delta_{1}A_{0}-\mu I_{0}-dI_{0}.\label{eq:6-2-1}
\end{equation}
Solving equations (\ref{eq:4-1-1}), (\ref{eq:5-1-2}) and (\ref{eq:6-2-1})
for $\beta_{N}$ and $A_{0}$ lead to
\begin{equation}
\beta_{N}=\frac{\left({\displaystyle a}+\delta\right)\left({\displaystyle a}+\mu+d\right)}{{\displaystyle a}+\mu+d+\delta_{1}}e^{a\tau}\mbox{ and }I_{0}=\frac{\delta_{1}}{{\displaystyle a}+\mu+d}A_{0}.\label{eq:5-1-1-1}
\end{equation}
 Furthermore, the initial data $A_{0}(\theta)$ and $I_{0}(\theta)$
are given for $\theta\in[-\tau,0]$ by $A_{0}(\theta)={\displaystyle \frac{ba}{\delta_{2}}}e^{a\theta}$
and $I_{0}(\theta)={\displaystyle \frac{\delta_{1}}{a+\mu+d}}A_{0}(\theta).$
We will assume, as in Section \ref{sec:A-Basic-COVID-19}, that the
latency duration varies between $6$ and $12$ hours. Consequently,
$\beta_{N}$ is estimated to be between $0.4492$ and $0.48$ with
an average of $0.4596.$ In this case, the basic reproduction number
varies between $2.96$ and $3.174$ with $3.03$ in average. Finally,
by repeating the same above process between the first and last day
of lockdown we can estimate the parameters $\beta_{c}$ and $p$ (See
Table \textsf{\small{}\ref{parametervalues-1-2-1-1}}). Here, the
sum of squared residuals is estimated to be $SSR_{dde}=92.63.$ When
considering $6$ hours and $12$ hours as latency periods then their
$SSR$ are given respectively by $SSR_{\tau=0.25}=93.23$ and $SSR_{\tau=0.5}=95.55.$

\begin{table}[!t]
\caption{Parameter definitions and values of  model (\ref{eq:delay}).}
\textsf{\small{}\label{parametervalues-1-2-1-1}}%
\begin{tabular}{>{\raggedright}p{1.5cm}>{\raggedright}p{3.9cm}p{2cm}>{\raggedright}p{2cm}>{\raggedright}p{1.7cm}}
\hline
\noalign{\smallskip{}
} Symbol & Definition  & Parameter value & Confidence Interval ($95\%$) & Reference\tabularnewline
\noalign{\smallskip{}
}\svhline\noalign{\smallskip{}
} $\beta_{N}$ & Infection rate for unconfined population & $0.4596$ & $0.4492-0.48$ & Estimated\tabularnewline
\textbf{$\beta_{c}$} & Infection rate for confined population & $0.091$ & $0.089-0.096$ & Estimated\tabularnewline
$\tau$ & latency period & $0.33$ & $0.25-0.5$ & See text\tabularnewline
$R_{0}$ & Basic reproduction number & $3.03$ & $2.96-3.174$ & Estimated\tabularnewline
$p$ & Proportion of lockdown & $0.75$ & $0.53-0.82$ & Estimated\tabularnewline
\noalign{\smallskip{}
}\noalign{\smallskip{}
} &  &  &  & \tabularnewline
\end{tabular}
\end{table}

\section{COVID-19 Model with Threshold-Type Delay \label{sec:COVID-threshold--Model}}

Threshold delay equations (TDEs) ensue in a natural way in compartmental
models for which the time in residence in a particular compartment
is determined by the stipulation that a fixed threshold load of an
entity is racked up during the time spent in that compartment. A susceptible
individual that is first exposed to a pathogen at time $t-\sigma$
will become infectious at time $t$ provided the individual receives
a sufficient load of the virus during the time from $t-\sigma$ to
$t.$ We will assume that an individual is exposed to an infectious
quantum, $c$, which is the unit of SARS-CoV-2 viral load needed to
produce an infection. Therefore, we will assume that the infectious
SARS-CoV-2 viral load will grow, overcoming the non-specific immune
response. When the pathogen load has increased to a threshold $Q,$
or equivalently, when the age since exposure is greater than the latency
period $\tau,$ we then consider the individual to be infectious.
We assume, as mentioned in \cite{Heneghan}, that the repeated exposures
to smaller viral loads increase the pathogen load in-host. Furthermore,
since transmission occurs from infected individuals, the pathogen
load due to an exposure will depend on the infected population.

Let $r$ be the internal growth rate of the SARS-CoV-2 virus, $b$
is the number of effective contacts between an exposed and infectious
individuals, $k$ is an adjustable parameter which measures how soon
saturation occurs. Following the modeling approach in \cite{Qesmi},
the threshold condition is governed by the following formula
\begin{equation}
\Psi(t):={\displaystyle ce^{r\tau(A_{t}+I_{u,t})}+\int_{-\tau(A_{t}+I_{u,t})}^{0}e^{-vr}G(A(t+v)+I_{u}(t+v))dv-Q=0}\label{eq:Threshold eq-1}
\end{equation}
where $A_{t}+I_{u,t}$ are the history functions of the infectious
individuals defined for $\xi\in[-\tau^{\infty}:=-\max_{\phi\in C}\tau(\phi),0]$
by
\[
A_{t}(\xi)+I_{u,t}(\xi)=A(t+\xi)+I_{u}(t+\xi)
\]
and $\tau:C\mapsto\mathbb{R}^{+}$ is a decreasing and continuously
differential map on the space of continuous functions,$C:=C\left([-\tau^{\infty},0],\mathbb{R}^{+}\right),$
satisfying $\tau(0)=\frac{1}{r}\ln\left(\frac{Q}{c}\right).$ Furthermore,
$F$ is the additive SARS-CoV-2 viral load in the exposed individual
due to multiple exposures to infectious individuals which is given,
for $x\geq0,$ by the following Holling functional response-type 2
\textbf{
\[
G(x)=\frac{bcx}{kx+1}.
\]
}The COVID-19 model will then be given by the following threshold-type
delay system

\begin{equation}
\begin{cases}
\frac{dS}{dt} & =-\alpha S(t)\left(A(t)+I_{u}(t)\right)/N,\vspace{0.05in}\\
{\displaystyle \frac{dA}{dt}} & =\alpha e^{-\eta\tau\left(A_{t}+I_{u,t}\right)}S\left(t-\tau\left(A_{t}+I_{u,t}\right)\right)\left(A\left(t-\tau\left(A_{t}+I_{u,t}\right)\right)+I_{u}\left(t-\tau\left(A_{t}+I_{u,t}\right)\right)\right)\\
 & -\delta A(t),\vspace{0.05in}\\
{\displaystyle \frac{dI_{u}}{dt}} & =\delta_{1}A(t)-\mu I_{u}(t)-dI_{u}(t),\vspace{0.05in}\\
\frac{dH}{dt} & =\delta_{2}A(t)-\mu H-dH,\vspace{0.05in}\\
{\displaystyle \frac{dR}{dt}} & =\delta B(t)+\mu(H+I_{u}),\vspace{0.05in}\\
{\displaystyle \frac{dD}{dt}} & =d(H+I_{u}),\vspace{0.05in}\\
\Psi(t) & =0
\end{cases}\label{eq:Threshold delay}
\end{equation}
where $\tau,$ $A$ and $I_{u}$ satisfy the threshold condition (\ref{eq:Threshold eq-1}).

Applying the survival function approach described by Heffernan, Smith,
and Wahl \cite{Heffernen2005} as done in Subsection \ref{subsec:Basic-reproduction-number delay},
the control and basic reproduction numbers are given by
\[
R_{c}=\frac{\alpha S_{0}e^{-\eta\tau\left(0\right)}\left(\mu+d+\delta_{1}\right)}{\delta N\left(\mu+d\right)}\mbox{ and }R_{0}=\frac{\beta_{N}S_{0}e^{-\eta\tau\left(0\right)}\left(\mu+d+\delta_{1}\right)}{\delta N\left(\mu+d\right)}.
\]
\begin{table}[t]
\caption{Parameter definitions and values of model (\ref{eq:Threshold delay}).}
\textsf{\small{}\label{parametervalues-1-2-1}}%
\begin{tabular}{>{\raggedright}p{1.5cm}>{\raggedright}p{3.9cm}p{2cm}>{\raggedright}p{2cm}>{\raggedright}p{1.7cm}}
\hline
\noalign{\smallskip{}
} Symbol & Definition & Parameter value & Confidence Interval ($95\%$) & Reference\tabularnewline
\noalign{\smallskip{}
}\svhline\noalign{\smallskip{}
} $\beta_{N}$ & Infection rate for unconfined population & $0.4596$ & $0.4492-0.48$ & Estimated\tabularnewline
$\beta_{c}$ & Infection rate for confined population & $0.091$ & $0.089-0.096$ & Estimated\tabularnewline
$r$ & Internal growth rate & $1.02$ &  & Adjusted\tabularnewline
$b$ & Effective contact number & $20$ &  & Adjusted\tabularnewline
$k$ & Adjustable parameter & $10^{-5}$ &  & Adjusted\tabularnewline
$c$ & Viral load per contact & $49,79$ & $18.31-135.33$ & Estimated\tabularnewline
$\tau(0)$ & Maximal latency duration & $0.33$ days & $0.25-0.25$ & See text\tabularnewline
$R_{0}$ & Basic reproduction number & $3.03$ & $2.96-3.174$ & Estimated\tabularnewline
$p$ & Proportion of lockdown & $0.73$ & $0.51-0.78$ & Estimated\tabularnewline
\end{tabular}
\end{table}
In order to estimate the model parameters and simulate its dynamics
we use MATLAB ddesd solver \cite{Shampine}\textbf{ }for state-dependent
delay differential equations to compute the solutions of (\ref{eq:Threshold delay})
numerically. However, we should note that simulating the behavior
of solutions of system (\ref{eq:Threshold delay}) for the general
state-dependent delay $\tau$ is a challenging task. To overcome this
difficulty we will estimate the parameters and perform our simulations
using constant initial data. Let $\tilde{C}=\left\{ \phi\in C:\phi(s)=\phi(0)\mbox{ for all }s\in[-\tau^{\infty},0]\right\} $
be the space of constant initial data.\textbf{ }Thus, for $\phi\in\tilde{C},$
the equation $\Psi(\tau(\phi),\phi)=0$ given by (\ref{eq:Threshold eq-1})
is equivalent to
\[
ce^{r\tau(\phi)}+\int_{-\tau(\phi)}^{0}e^{-rs}G(\phi(0))dv-Q=0.
\]
Solving this equation for $\tau(\phi)$, we obtain
\begin{equation}
\tau(\phi)=\frac{1}{r}\ln\left(\frac{rQ+G(\phi(0))}{cr+G(\phi(0))}\right).\label{eq:Sigma-sim}
\end{equation}
It is experimentally shown that the minimal viral load needed for
the infection to occur in hamsters is $1000$ particles (See \cite{Karimzadeh}
more details). Thus, we assume that $Q=1000.$ Furthermore, since
the maximal latency duration $\tau(0)$ varies between $6$ and $12$
hours then, from formula (\ref{eq:Sigma-sim}) the viral load, $c,$
per each contact will vary between $Qe^{-r/2}$ and $Qe^{-r/4}$ with
an average of $Qe^{-r/3}.$ However, another difficulty we encounter
for this model is that the parameters $b,$ $r$ and $k$ related
to COVID-19 disease are still unknown and we are compelled to fairly
adjust them to fit the reported cases. (See Table \ref{parametervalues-1-2-1}).
In this case we will be able to follow the same process as the one
in Subsection \ref{subsec:Parameters-and-initial} to obtain the remaining
model parameters (See Table \ref{parametervalues-1-2-1}). Furthermore
we obtain $SSR_{sde}=94.3.$

\section{Models with Demographic Effects \label{sec:Models-with-Demographic}}

In the previous section we have omitted births and deaths in our description
of models because it was believed that the time scale of of COVID-19
epidemic is much shorter than the demographic time scale. Indeed,
we have used a time scale on which the number of births and deaths
in unit time is negligible. However, as mentioned in Section \ref{sec:Introduction},
there is a possibility that the COVID-19 may not go away after a short
time and could stay for years. Thus, we need to think on a longer
time scale and include a birth rate parameter, $\pi_{S}$ and a death
rate parameter $d_{S}.$ In what follows, we will reconsider models
(\ref{eq:delay}) and (\ref{eq:Threshold delay}) including demographic
effects and we shall give a rigorous mathematical analysis to the
both models. The reason for which we select these models is that model
(\ref{eq:delay}) is shown to be the best one to fit well the data
while model (\ref{eq:Threshold delay}), as we will see in Subsection
\ref{subsec:Threshold-Math}, generates more complicated behavior
then the three other models.

\subsection{COVID-19 Model with Constant Delay \label{subsec:Constant delay2}}

Let us analyze the following constant delay COVID-19 model

\begin{equation}
\begin{cases}
\frac{dS}{dt} & =\pi_{S}-\alpha e^{-\eta\tau}S(t)\left(A(t)+I_{u}(t)\right)/N-d_{S}S(t),\vspace{0.05in}\\
{\displaystyle \frac{dA}{dt}} & =\alpha e^{-\eta_{a}\tau}S(t-\tau)\left(A(t-\tau)+I_{u}(t-\tau)\right)-\delta A(t)--d_{S}A(t),\vspace{0.05in}\\
{\displaystyle \frac{dI_{u}}{dt}} & =\delta_{1}A(t)-\mu I_{u}(t)-dI_{u}(t)-d_{S}I_{u}(t),\vspace{0.05in}\\
\frac{dH}{dt} & =\delta_{2}A(t)-\mu H(t)-dH(t)-d_{S}H(t),\vspace{0.05in}\\
{\displaystyle \frac{dR}{dt}} & =\mu\left(H(t)+I_{u}(t)\right)-d_{S}R(t),\vspace{0.05in}\\
{\displaystyle \frac{dD}{dt}} & =d\left(H(t)+I_{u}(t)\right)
\end{cases}\label{eq:Delay2}
\end{equation}
where
\[
\alpha=\left(\left(1-\gamma\right)p\beta_{c}+(1-\theta)\left(1-\left(1-\gamma\right)p\right)\beta_{N}\right).
\]
Since the three last components $H,$$R$ and $D$ do not appear in
the three first equations of model (\ref{eq:Delay2}) then we will
focus our local stability study on the three first equations.

\subsubsection{Equilibria}

Computing the equilibria of system (\ref{eq:Delay2}) we see that
a positive steady state $\left(\tilde{S},\tilde{A},\tilde{I}_{u}\right)$
must satisfy

\[
\begin{cases}
\pi_{S}-\alpha e^{-\eta\tau}\tilde{S}\left(\tilde{A}+\tilde{I}_{u}\right)/N-d_{S}\tilde{S} & =0,\vspace{0.05in}\\
\alpha e^{-\eta\tau}\tilde{S}\left(\tilde{A}+\tilde{I}_{u}\right)/N-\left(\delta+d_{S}\right)\tilde{A} & =0,\vspace{0.05in}\\
\delta_{1}\tilde{A}-\left(\mu+d+d_{S}\right)\tilde{I}_{u} & =0.
\end{cases}
\]
A straightforward calculation of the above system leads to the following
result.
\begin{proposition}
The model (\ref{eq:Delay2}) has a disease free equilibrium (DFE)
given by $\overline{E}=(\frac{\pi_{S}}{d_{S}},0,0)$ in which there
is no disease. Furthermore, the DFE is unique when $R_{c}\leq1$ and
a unique endemic equilibrium of (\ref{eq:Delay2}) appears when $R_{c}>1.$

\noindent Moreover, the endemic equilibrium, $E^{*}=\left(\tilde{S},\tilde{A},\tilde{I}_{u}\right),$
satisfies
\end{proposition}
\begin{equation}
\tilde{S}=\frac{\delta}{\alpha e^{-\eta\tau}\left(1+\frac{\delta_{1}}{\mu+d+d_{S}}\right)/N},\tilde{A}=\delta\left(\pi_{S}-d_{S}\tilde{S}\right)/N\mbox{ and }\tilde{I}_{u}=\frac{\delta_{1}}{\mu+d+d_{S}}\tilde{A},\label{Nstar-ss}
\end{equation}

\subsubsection{Control Thresholds}

Following the work done in Subsection \ref{subsec:Basic-reproduction-number delay}
we can easily obtain the basic reproduction number related to model
(\ref{eq:Delay2}) as follows

\[
R_{0}\mbox{=}\frac{\pi_{S}\beta_{N}e^{-\eta\tau}}{d_{S}N}\left(\frac{1}{\delta+d_{S}}+\frac{\delta_{1}}{\left(\delta+d_{S}\right)\left(\mu+d+d_{S}\right)}\right).
\]
The control reproduction number is given by
\[
R_{c}\mbox{=}\frac{\pi_{S}\alpha e^{-\eta\tau}}{d_{S}N}\left(\frac{1}{\delta+d_{S}}+\frac{\delta_{1}}{\left(\delta+d_{S}\right)\left(\mu+d+d_{S}\right)}\right).
\]

\subsubsection{Bifurcation Analysis}

In the following,\textcolor{blue}{{} }we shall prove that system (\ref{eq:Delay2})
produces a forward transcritical bifurcation. The linearization of
system (\ref{eq:Delay2}) around any steady state E = $\left(S^{e},A^{e},I_{u}^{e}\right)$
characteristic equation
\begin{eqnarray}
\Delta\left(\lambda\right) & = & \left(\lambda+\delta+d_{S}\right)\left(\lambda+\alpha e^{-\eta\tau}\left(A^{e}+I_{u}^{e}\right)/N+d_{S}\right)\left({\displaystyle \lambda}+\mu+d+d_{S}\right)\label{eq:Charact}\\
 &  & -\alpha e^{-\eta\tau}S^{e}\left({\displaystyle \lambda}+\mu+d+\delta_{1}+d_{S}\right)\left(\lambda+d_{S}\right)e^{-\lambda\tau}/N.\nonumber
\end{eqnarray}
The local behavior of the DFE of system (\ref{eq:Delay2}) is given
by the following theorem.
\begin{theorem}
\label{thm:The-boundary-steady} The boundary steady state $\overline{E}$
of system (\ref{eq:Delay2}) is unstable when $R_{c}>1$ and locally
asymptotically stable when $R_{c}<1.$
\end{theorem}
\begin{proof}
The characteristic equation associated with the DFE is given by
\[
\Delta(\lambda)=\left(\lambda+d_{S}\right)\left(\left(\lambda+\delta+d_{S}\right)\left({\displaystyle \lambda}+\mu+d+d_{S}\right)-\alpha e^{-\eta\tau}\overline{S}\left({\displaystyle \lambda}+\mu+d+\delta_{1}+d_{S}\right)e^{-\lambda\tau}/N\right).
\]
Then the associated eigenvalues are given by $\lambda=-d_{S}$ and
the roots of
\begin{equation}
\tilde{\delta}\left(\lambda\right)=\left(\lambda+\delta+d_{S}\right)\left({\displaystyle \lambda}+\mu+d+d_{S}\right)-\alpha e^{-\eta\tau}\overline{S}\left({\displaystyle \lambda}+\mu+d+\delta_{1}+d_{S}\right)e^{-\lambda\tau}/N.\label{eq:characDFE1}
\end{equation}
Let $\lambda$ be any eigenvalue associated to equation (\ref{eq:characDFE1})
with nonnegative real part (i.e. $\Re e\left(\lambda\right)\geq0$)
and assume that $R_{c}<1.$ Then $e^{-\Re e\left(\lambda\right)\tau}\leq1$
and
\[
\frac{\mid\lambda+\delta+d_{S}\mid\mid{\displaystyle \lambda}+\mu+d+d_{S}\mid}{\mid{\displaystyle \lambda}+\mu+d+\delta_{1}+d_{S}\mid}\leq\alpha e^{-\eta\tau}\overline{S}/N.
\]
 On the other hand, it follows from $R_{c}<1$ that
\[
\alpha e^{-\eta\tau}\overline{S}/N<\frac{\left(\delta+d_{S}\right)\left(\mu+d+d_{S}\right)}{\mu+d+\delta_{1}+d_{S}}.
\]

This is a contradiction since the map $\lambda\mapsto{\displaystyle \frac{\mid\lambda+\delta+d_{S}\mid\mid{\displaystyle \lambda}+\mu+d+d_{S}\mid}{\mid{\displaystyle \lambda}+\mu+d+\delta_{1}+d_{S}\mid}}$
is increasing. Consequently, the DFE is locally asymptotically stable.
On the other hand if $R_{c}>1$ then
\[
\tilde{\delta}\left(0\right)=\delta\left(\mu+d+d_{S}\right)\left(1-R_{c}\right)<0.
\]
Then $\tilde{\delta}$ has one positive root and the DFE is unstable.
\end{proof}

The forward transcritical bifurcation of the endemic equilibrium as
$R_{c}$ moves through $1$ is stated as follows.
\begin{theorem}
When $R_{c}<1,$ the endemic equilibrium of system (\ref{eq:Delay2})
is locally asymptotically stable while the DFE is unstable, and for
$R_{c}>1$ the DFE is unique and locally asymptotically stable. That
is, forward transcritical bifurcation occurs at $R_{c}=1.$
\end{theorem}
\begin{proof}
The characteristic equation associated with the endemic equilibrium
is given by
\[
\Delta\left(\lambda\right)=\frac{\left(\lambda+\delta+d_{S}\right)\left(\lambda+\alpha e^{-\eta\tau}\left(\tilde{A}+\tilde{I}_{u}\right)/N+d_{S}\right)\left({\displaystyle \lambda}+\mu+d+d_{S}\right)}{\left({\displaystyle \lambda}+\mu+d+\delta_{1}+d_{S}\right)\left(\lambda+d_{S}\right)}-\alpha e^{-\eta_{a}\tau}\tilde{S}/Ne^{-\lambda\tau}.
\]
Let $\lambda$ be any eigenvalue associated to equation (\ref{eq:characDFE1})
with nonnegative real part. Then
\begin{align*}
\frac{\mid\lambda+\delta+d_{S}\mid\mid{\displaystyle \lambda}+\mu+d+d_{S}\mid}{\mid{\displaystyle \lambda}+\mu+d+\delta_{1}+d_{S}\mid} & \leq\frac{\mid\lambda+\delta+d_{S}\mid\mid\lambda+\alpha e^{-\eta\tau}\left(\tilde{A}+\tilde{I}_{u}\right)/N+d_{S}\mid\mid{\displaystyle \lambda}+\mu+d+d_{S}\mid}{\mid{\displaystyle \lambda}+\mu+d+\delta_{1}+d_{S}\mid\mid\lambda+d_{S}\mid}\\
 & =\alpha e^{-\eta\tau}\tilde{S}\mid e^{-\lambda\tau}\mid/N\\
 & \leq\frac{\left(\mu+d+d_{S}\right)\delta}{\mu+d+\delta_{1}+d_{S}}
\end{align*}
 However, ${\displaystyle \lambda\mapsto\frac{\mid\lambda+\delta+d_{S}\mid\mid{\displaystyle \lambda}+\mu+d+d_{S}\mid}{\mid{\displaystyle \lambda}+\mu+d+\delta_{1}+d_{S}\mid}}$
is increasing which is a contradiction. It follows that all characteristic
roots of $\Delta$ are negative. Thus, the local asymptotic stability
of the positive steady state immediately follows. Furthermore, from
Theorem \ref{thm:The-boundary-steady} we deduce the local behavior
of the DFE. This completes the proof.
\end{proof}

\subsection{COVID-19 Model with Threshold-Type Delay \label{subsec:Threshold-Math}}

In this section we will perform qualitative analysis of the following
threshold-type delay COVID-19 model. We shall prove that system (\ref{eq:Delay2})
produces two potential cases of bifurcation depending on the chosen
parameter values.

\begin{equation}
\begin{cases}
\frac{dS}{dt} & =\pi_{S}-\alpha S(t)\left(A(t)+I_{u}(t)\right)/N-d_{S}S,\vspace{0.05in}\\
{\displaystyle \frac{dA}{dt}} & =\alpha e^{-\eta\tau\left(A(t)+I_{u}(t)\right)}S(t-\sigma(t))\left(A(t-\sigma(t))+I_{u}(t-\sigma(t))\right)\\
 & -\left(\delta+d_{S}\right)A(t),\vspace{0.05in}\\
{\displaystyle \frac{dI_{u}}{dt}} & =\delta_{1}A(t)-\mu I_{u}(t)-\left(d+d_{S}\right)I_{u}(t),\vspace{0.05in}\\
\frac{dH}{dt} & =\delta_{2}A(t)-\mu H-\left(d+d_{S}\right)H,\vspace{0.05in}\\
{\displaystyle \frac{d\tilde{R}}{dt}} & =\delta B(t)+\mu(H+I_{u}),\vspace{0.05in}\\
{\displaystyle \frac{dD}{dt}} & =d(H+I_{u})\vspace{0.05in}
\end{cases}\label{eq:Threshold2}
\end{equation}
 where $\sigma(t)=\tau\left(A(t)+I_{u}(t)\right).$

\subsubsection{Equilibria}

As mentioned in Subsection \ref{subsec:Constant delay2} we will focus
our study only on the three first equations. Computing the equilibria
of the system (\ref{eq:Threshold2}) we see that an endemic equilibrium$\left(\tilde{S},\tilde{A},\tilde{I}_{u}\right)$
must satisfy

\[
\begin{cases}
\pi_{S}-\alpha\tilde{S}\left(\tilde{A}+\tilde{I}_{u}\right)/N-d_{S}\tilde{S} & =0,\vspace{0.05in}\\
\alpha e^{-\eta\tau\left(\tilde{A}+\tilde{I}_{u}\right)}\tilde{S}\left(\tilde{A}+\tilde{I}_{u}\right)/N-\left(\delta+d_{S}\right)\tilde{A} & =0,\vspace{0.05in}\\
\delta_{1}\tilde{A}-\left(\mu+d+d_{S}\right)\tilde{I}_{u} & =0.
\end{cases}
\]
Let, for $y>0,$ $\nu\left(y\right)=\tau\left(\left({\displaystyle \frac{\mu+d+\delta_{1}+d_{S}}{\mu+d+d_{S}}}\right)y\right),$
\[
W(y)=\eta\nu'\left(y\right)\left(\alpha\left({\displaystyle \frac{\mu+d+\delta_{1}+d_{S}}{\mu+d+d_{S}}}\right)y/N+d_{S}\right)+\alpha\left({\displaystyle \frac{\mu+d+\delta_{1}+d_{S}}{\mu+d+d_{S}}}\right)/N
\]
 and
\[
\chi\left(y\right)=\frac{\pi_{S}\alpha e^{-\eta\nu\left(y\right)}\left(\mu+d+\delta_{1}+d_{S}\right)}{\alpha\left({\displaystyle \frac{\mu+d+\delta_{1}+d_{S}}{\mu+d+d_{S}}}\right)y/N+d_{S}}.
\]
 A straightforward calculation of the above system leads to the following
result.
\begin{proposition}
\label{prop:Equilib} The model (\ref{eq:Threshold2}) has a disease
free equilibrium (DFE) given by,
\[
\overline{E}=(\frac{\pi_{S}}{d_{S}},0,0)
\]
in which there is no disease. Furthermore,

(i) if $R_{c}\leq1$ and $W\left(\tilde{A}\right)>0$ then there is
no endemic equilibria,

(ii ) if $R_{c}>1$ and $W\left(\tilde{A}\right)>0$ then there is
exists only one endemic equilibrium,

(iii) if $R_{c}<1$ and there exist $A^{*}>0$ such that $\chi(A^{*})>\delta N\left(\mu+d\right).$
then there exist at least two endemic equilibria,

(vi) if $R_{c}\leq1$ and, for all $y>0,$ $\chi(y)<\delta N\left(\mu+d\right).$
then there is no endemic equilibria.

\noindent Moreover, the endemic equilibrium, $E^{*}=\left(\tilde{S},\tilde{A},\tilde{I}_{u}\right),$
satisfies
\end{proposition}
\begin{equation}
\tilde{S}=\frac{\pi_{S}}{\alpha\left({\displaystyle \frac{\mu+d+\delta_{1}+d_{S}}{\mu+d+d_{S}}}\right)\tilde{A}/N+d_{S}},\tilde{I}_{u}=\frac{\delta_{1}}{\mu+d+d_{S}}\tilde{A}\mbox{ and }\chi\left(\tilde{A}\right)=\delta N\left(\mu+d+d_{S}\right).\label{Nstar-ss-1}
\end{equation}

\begin{proof}
After few calculation we obtain
\[
\chi\left(0\right)={\displaystyle \frac{\pi_{S}\alpha e^{-\eta\nu\left(0\right)}\left(\mu+d+\delta_{1}+d_{S}\right)}{d_{S}}=R_{c}\delta N\left(\mu+d+d_{S}\right)},
\]
 $\lim_{y\rightarrow\infty}\chi(y)=0$ and
\[
\chi'\left(\tilde{A}\right)=-\frac{\pi_{S}\alpha\left(\mu+d+\delta_{1}+d_{S}\right)e^{-\eta\nu\left(\tilde{A}\right)}}{\left(\alpha\left({\displaystyle \frac{\mu+d+\delta_{1}+d_{S}}{\mu+d+d_{S}}}\right)\tilde{A}/N+d_{S}\right)^{2}}W\left(\tilde{A}\right).
\]
This proves all the assertions of the proposition.
\end{proof}

\subsubsection{Control Thresholds}

Similarly to the proof in Subsection \ref{subsec:Basic-reproduction-number delay},
the basic and control reproduction numbers for system (\ref{eq:Threshold2})
are successively given by
\[
R_{0}=\frac{\pi_{S}\beta_{N}e^{-\eta\nu\left(0\right)}\left(\mu+d+\delta_{1}+d_{S}\right)}{\delta N\left(\mu+d+d_{S}\right)d_{S}}\mbox{ and }R_{c}\mbox{=}\frac{\pi_{S}\alpha e^{-\eta\nu\left(0\right)}}{Nd_{S}}\left(\frac{1}{\delta+d_{S}}+\frac{\delta_{1}}{\left(\delta+d_{S}\right)\left(\mu+d+d_{S}\right)}\right).
\]

\subsubsection{Bifurcations}

Here we focus on local asymptotic stability and bifurcation analysis
of equilibria of system (\ref{eq:Threshold2}) .
\begin{theorem}
\label{thmstabtss} The DFE $\overline{E}=\left(\overline{S},0,0\right)$
of (\ref{eq:Threshold2}) is unstable when ${\displaystyle R_{c}>1}$,
and locally asymptotically stable when ${\displaystyle R_{c}<1.}$
\end{theorem}
\begin{proof}
The characteristic equation associated with the DFE is given by
\begin{eqnarray*}
\Delta\left(\lambda\right) & = & \left(\lambda+\delta+d_{S}\right)\left(\lambda+\mu+d+d_{S}\right)-\alpha e^{-\eta\nu\left(0\right)}\overline{S}e^{-\lambda\nu\left(0\right)}\left(\lambda+\mu+d+\delta_{1}+d_{S}\right)/N\\
 & = & \lambda^{2}+\left(\mu+d+\delta+d_{S}-\alpha\overline{S}/N\right)\lambda-\alpha e^{-\eta\nu\left(0\right)}\overline{S}/N\left(\lambda+\mu+d+\delta_{1}+d_{S}\right)e^{-\lambda\nu\left(0\right)}\\
 &  & +\left(\delta+d_{S}\right)\left(\mu+d+d_{S}\right).
\end{eqnarray*}
 When $\nu=0$ then $\Delta\left(\lambda\right)=\lambda^{2}+\left(\mu+d+\delta+d_{S}-\alpha\overline{S}/N\right)\lambda+\delta\left(\mu+d+d_{S}\right)\left(1-R_{c}\right).$
Furthermore, we have
\begin{align*}
\mu+d+\delta+d_{S}-\overline{S}\alpha/N & =\frac{\left(\delta+d_{S}\right)\left(\mu+d+d_{S}\right)}{\mu+d+\delta_{1}+d_{S}}\frac{\left(\mu+d+\delta+d_{S}\right)\left(\mu+d+\delta_{1}+d_{S}\right)}{\left(\delta+d_{S}\right)\left(\mu+d+d_{S}\right)}\\
 & >\frac{\left(\delta+d_{S}\right)\left(\mu+d+d_{S}\right)}{\mu+d+\delta_{1}+d_{S}}\left(1-R_{c}\right).
\end{align*}
 Thus, the DFE is stable when $\nu=0$ and $R_{c}<1.$ Assume that\textbf{
$\nu>0$ }and let $\omega>0.$ Separating real and imaginary parts,
equality $\Delta(i\omega)=0$ is equivalent to
\begin{eqnarray*}
\omega^{4}+\left(\left(\mu+d+d_{S}\right)^{2}+\left(\delta+d_{S}\right)^{2}-\left(\alpha e^{-\eta\nu\left(0\right)}\overline{S}/N\right)^{2}\right)\omega^{2}+\left(\delta+d_{S}\right)^{2}\left(\mu+d+d_{S}\right)^{2}\\
-\left(\alpha e^{-\eta\nu\left(0\right)}\overline{S}/N\left(\mu+d+\delta_{1}+d_{S}\right)\right)^{2}=0.
\end{eqnarray*}
 A simple calculation of the discriminant $\tilde{\delta}$ leads
to
\begin{eqnarray*}
\tilde{\delta} & = & \left(\left(\mu+d+d_{S}\right)^{2}-\left(\delta+d_{S}\right)^{2}-\left(\alpha e^{-\eta\nu\left(0\right)}\overline{S}/N\right)^{2}\right)^{2}+4\left(\delta+d_{S}\right)^{2}\left(\alpha e^{-\eta\nu\left(0\right)}\overline{S}/N\right)^{2}\\
 &  & +4\left(\alpha e^{-\eta\nu\left(0\right)}\tilde{S}/N\left(\mu+d+\delta_{1}+d_{S}\right)\right)^{2}
\end{eqnarray*}
 which is positive. It follows that $i\omega$ is not a root of $\Delta$
and, consequently, the DFE is LAS for all $\nu>0$ such that $R_{c}<1.$
On the other hand, if $R_{c}>1$ then
\begin{eqnarray*}
\Delta(0) & = & \delta\left(\mu+d+d_{S}\right)\left(1-\frac{\alpha e^{-\eta\nu\left(0\right)}\pi_{S}\left(\mu+d+\delta_{1}+d_{S}\right)/N}{d_{S}\delta\left(\mu+d+d_{S}\right)}\right)\\
 &  & =\left(\delta+d_{S}\right)\left(\mu+d+d_{S}\right)\left(1-R_{c}\right)
\end{eqnarray*}
 which is negative. It follows that $\Delta$ has a positive root
and the DFE is unstable.
\end{proof}

\begin{theorem}
\label{thmtransbif-1} When $R_{c}=1$ and $W\left(\tilde{A}\right)>0$
, the endemic equilibrium undergoes a forward transcritical bifurcation,
that is for $R_{c}>1$, $R_{c}$ close to $1$, the endemic equilibrium
is locally asymptotically stable whereas the DFE is unstable, and
for $R_{c}<1$ the DFE is locally asymptotically stable and is the
only steady state of (\ref{eq:Threshold2}).
\end{theorem}
\begin{proof}
The characteristic equation is given by
\begin{align*}
\Delta\left(\lambda\right) & =\left(\lambda+\delta+d_{S}\right)\left(\lambda+\mu+d+d_{S}\right)-\alpha\tilde{S}e^{-\eta\nu\left(\tilde{A}\right)}e^{-\lambda\nu\left(\tilde{A}\right)}\left(\lambda+\mu+d+\delta_{1}+d_{S}\right)/N\\
 & +Q(\lambda)
\end{align*}
where
\[
Q\left(\lambda\right)=\alpha e^{-\eta\nu\left(\tilde{A}\right)}\left(\tilde{A}+\tilde{I}_{u}\right)\tilde{S}\left(\lambda+\mu+d+\delta_{1}+d_{S}\right)\left(\frac{e^{-\lambda\nu\left(\tilde{A}\right)}\alpha/N}{\lambda+\alpha\left(\tilde{A}+\tilde{I}_{u}\right)/N+d_{S}}+\eta\nu'\left(\tilde{A}\right)\right)/N.
\]
Then
\begin{eqnarray*}
\Delta\left(0\right) & = & \delta\left(\mu+d+d_{S}\right)-\alpha e^{-\eta\nu\left(\tilde{A}\right)}\tilde{S}\left(\mu+d+\delta_{1}+d_{S}\right)/N\\
 &  & +\alpha e^{-\eta\nu\left(\tilde{A}\right)}\left(\tilde{A}+\tilde{I}_{u}\right)\tilde{S}\left(\mu+d+d_{S}\right)\left(\frac{\alpha\left(\frac{\mu+d+\delta_{1}}{\mu+d}\right)/N}{\alpha\left(\tilde{A}+\tilde{I}_{u}\right)/N+d_{S}}+\eta\nu'\left(\tilde{A}\right)\right)/N.
\end{eqnarray*}
On the other hand, from the equilibrium equation (\ref{Nstar-ss-1}),
we have
\[
\alpha e^{-\eta_{a}\nu\left(\tilde{A}\right)}\tilde{S}\left(\mu+d+\delta_{1}+d_{S}\right)/N-\left(\delta+d_{S}\right)\left(\mu+d+d_{S}\right)=0.
\]
Then
\[
\Delta\left(0\right)=\alpha e^{-\eta\nu\left(\tilde{A}\right)}\left(\tilde{A}+\tilde{I}_{u}\right)\tilde{S}\left(\mu+d+d_{S}\right)\left(\frac{\alpha\left({\displaystyle \frac{\mu+d+\delta_{1}+d_{S}}{\mu+d+d_{S}}}\right)/N}{\alpha\left(\tilde{A}+\tilde{I}_{u}\right)/N+d_{S}}+\eta\nu'\left(\tilde{A}\right)\right)/N.
\]
It follows from condition (ii) in Proposition \ref{prop:Equilib}
that $\Delta(0)>0.$ This proves that $\lambda=0$ is not a root of
$\Delta\left(\lambda\right)=0.$

Now, let $\lambda$ be a root of $\Delta(.)$ with nonnegative real
part and $Z=\mid\lambda+\mu+d+\delta_{1}+d_{S}\mid\alpha e^{-\eta\nu\left(\tilde{A}\right)}\tilde{S}/N.$
Thus, $\mid e^{-\lambda\nu\left(\tilde{A}\right)}\mid\leq1$ and
\begin{eqnarray*}
\mid Q\left(\lambda\right)\mid & \leq & Z\left(\frac{\alpha\left(\tilde{A}+\tilde{I}_{u}\right)/N}{\mid\lambda+\alpha\left(\tilde{A}+\tilde{I}_{u}\right)/N+d_{S}\mid}\right)\\
 & < & Z.
\end{eqnarray*}
Set $X=\frac{\left(\delta+d_{S}\right)\left(\mu+d+d_{S}\right)}{\mu+d+\delta_{1}+d_{S}}-\alpha\tilde{S}e^{-\eta\nu\left(\tilde{A}\right)}e^{-\lambda\nu\left(\tilde{A}\right)}/N$
and $Y=\alpha e^{-\eta_{a}\nu\left(\tilde{A}\right)}\tilde{S}/N-\alpha\tilde{S}e^{-\eta\nu\left(\tilde{A}\right)}e^{-\lambda\nu\left(\tilde{A}\right)}/N.$
Then, using formula \ref{Nstar-ss-1}, we have $X=Y$ and
\begin{align*}
\mid\left(\lambda+\delta+d_{S}\right)\left(\lambda+\mu+d+d_{S}\right)-\alpha\tilde{S}e^{-\eta\nu\left(\tilde{A}\right)}e^{-\lambda\nu\left(\tilde{A}\right)}/N\mid & \geq\mid\lambda+\mu+d+\delta_{1}+d_{S}\mid\mid Y\mid\\
 & =Z\mid1-e^{-\lambda\nu\left(\tilde{A}\right)}\mid.
\end{align*}
Therefore, $\mid1-e^{-\lambda\nu\left(\tilde{A}\right)}\mid<1$ which
is a contradiction since $\Re e(\lambda)\geq0.$ Consequently, the
endemic equilibrium is LAS.
\end{proof}

\begin{theorem}
\noindent \label{thm:backwbif} Assume that case (iii) in Proposition
\ref{prop:Equilib} holds true. When $R_{c}=1,$ the system (\ref{eq:Threshold2})
undergoes a backward bifurcation. That is, for $R_{c}>1$, $R_{c}$
close to $1$, the endemic equilibrium is the unique equilibrium which
is locally asymptotically stable; and for $R_{c}<1$ $R_{c}$ close
to $1$, the DFE together with an endemic equilibria are locally asymptotically
stable whereas a second endemic equilibrium exists and is unstable.
\end{theorem}
\begin{proof}
When case (iii) hold, then there exists at least two positive steady
states, $E_{m}=\left(S^{m},A^{m},I_{u}^{m}\right)$ and $E_{M}=\left(S^{M},A^{M},I_{u}^{M}\right).$
The selected equilibria $E_{m}$ and $E_{M}$ we will use are the
first two solutions $A^{m}$ and $A^{M}$ of equation $\chi(y)=\delta N\left(\mu+d\right)$such
that $\chi'(A^{M})<0$ and $\chi'(A^{m})>0.$ Thus, the proof of the
LAS of equilibrium $E^{M}$ is similar to the one of Theorem \ref{thmtransbif-1}.

The characteristic equation associated to $E_{m}$ satisfies
\[
\Delta\left(0\right)=\alpha e^{-\eta\nu\left(A_{m}\left(\zeta\right)\right)}\left(A_{m}\left(\zeta\right)+I_{m}\left(\zeta\right)\right)S_{m}\left(\zeta\right)\left(\mu+d+d_{S}\right)\chi'(A^{m})/N.
\]
Consequently $\Delta(0)<0$ and, since $\lim_{\lambda\rightarrow\infty}\Delta(\lambda)=+\infty,$
then there exists $\lambda^{*}>0$ such that $\Delta(\lambda^{*})=0$.
This concludes the proof of the theorem.
\end{proof}

The existence of a backward bifurcation is an interesting artifact
since this means that repeated exposures of susceptibles to the SARS-CoV-2
virus can cause bi-stability dynamics and, subsequently, infection
persistence even when the control reproduction number $R_{c}$ is
less than unity. An interesting query that emanates from the backward
bifurcation is \textquotedblleft What is the maximum effective contact
number, $b,$ or viral load per each contact, $c,$ below which the
COVID-19 disappear one we reduce $R_{c}$ below one?\textquotedblright .
Note that, as mentioned and proved in \cite{Qesmi}, in the case of
single exposure model (\ref{eq:Threshold2}) is a system of constant-delay
differential equations which is equivalent to system (\ref{eq:Delay2}).
Moreover, no backward bifurcation occurs. Generally, there is a threshold
$b^{*}$ below which the backward bifurcation disappear (The proof
is similar to the one in \cite{Qesmi}). This result could have a
significant biological interpretation since, as stated in \cite{Karimzadeh},
minimization of exposure to SARS-COV-2 is key to reducing the chance
of infection and developing disease.

\section{Discussion\label{sec:Lifting-Control}}

\begin{figure}[H]
\sidecaption 
\includegraphics[scale=0.28]{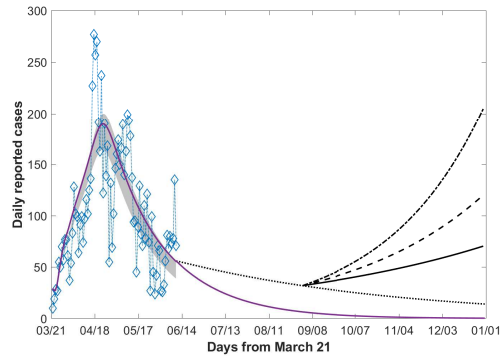}
\caption{Time series plot for model (\ref{eq:Threshold delay}), starting from
March $21,2020,$ of the numbers of reported individuals change using
different lifting rates at different times as follows: $30\%$ of
the total population is lifted on June $10$ (dot line) and a fraction
$\gamma$ of the remaining confined individuals ($\gamma=0.3,$ $\gamma=0.4,$
and $\gamma=0.5,$ solid line, dashed line, dot-dashed line respectively)
is lifted on September $1st.$}
\label{fig:SDE} 
\end{figure}

\begin{figure}[H]
\sidecaption 
\includegraphics[scale=0.28]{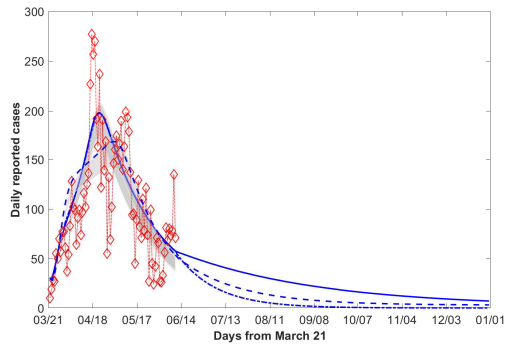}
\caption{Time series plot for model (\ref{eq:delay}), starting from March
$21,2020,$ of the numbers of reported individuals change using different
latent periods ($\tau=0.25,$ $\tau=0.33,$ and $\tau=0.5,$ dashed
line, solid line, dot-dashed line respectively) with $30\%$ of the
total population is lifted on June $10.$}
\label{fig:Effect} 
\end{figure}

Since the beginning of COVID-19 pandemic, numerous mathematical models
with increasing complexity are developed worldwide to understand the
course of COVID-19 disease. The modeling results have shown a wide
large of variations, especially in the basic reproduction numbers.
This lead to ask some questions such as: Why these variations exist
between models? Which model is the most realistic for the COVID-19
disease? WHO had reported that the basic reproduction number is estimated
to be between $1.4$ and $2.5$ \cite{Statement} while other interesting
contributions reported that the COVID-19 is more transmissible than
what WHO mentioned. In \cite{LiuSars}, the authors estimated, through
a comparison study of $12$ different results, that the median value
of $R_{0}$ for COVID-19 is expected to be around $2-3$. However,
only $6$ among these studies have used mathematical models leading
to a higher variation of $R_{0}$ $(1.5-6.49),$ with an average of
$4.2.$ Another systematic review in \cite{Lin} screened $75$ mathematical
and statistical models published between December $1st$ $2019$ and
February $21st$ $2020$ and concluded that the median of $R_{0}$
for COVID-19 was $3.77$.
\begin{figure}[H]
\sidecaption 
\includegraphics[scale=0.28]{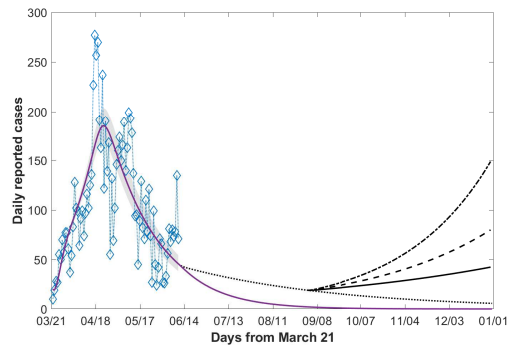}
\caption{Time series plot for model (\ref{eq:ODE}), starting from March $21,2020,$
of the numbers of reported individuals change using different lifting
rates at different times as follows: $30\%$ of the total population
is lifted on June $10$ (dot line) and a fraction $\gamma$ of the
remaining confined individuals ($\gamma=0.3,$ $\gamma=0.4,$ and
$\gamma=0.5,$ solid line, dashed line and dot-dashed line respectively)
is lifted on September $1st.$}
\label{fig:ODE} 
\end{figure}
 In this chapter, we developed, fitted and compared four mathematical
models with increasing complexity, that incorporate lifting lockdown
strategy, to check out which one among them provides the best prediction
for COVID-19 disease. We considered a progressive relaxation of the
compulsory lockdown performed in two stages and supported by a reduction
of 60\% of the contact rate. In the first stage 30 \% of the total
confined population lifted the lockdown on June 10th while, in the
second stage another 30\% lifted the lockdown on September 1st. Our
investigation of the proposed models showed a small variation of $R_{0}$
ranging from 2.06 to 3.03, and according to the $SSR$ measure (see
Section \ref{sec:A-Basic-COVID-19}), the best fit of reported data
is achieved for the model with constant delay (\ref{eq:delay}) with
$SSR=92.63.$ Consequently, our results show that model (\ref{eq:delay})
is the most reliable to estimate the value of $R_{0}$ ($R_{0}=3.03$),
which is higher than those estimated by models (\ref{eq:ODE}), (\ref{eq:Age})
and (\ref{eq:Threshold delay}).
\begin{figure}[H]
\sidecaption 
\includegraphics[scale=0.28]{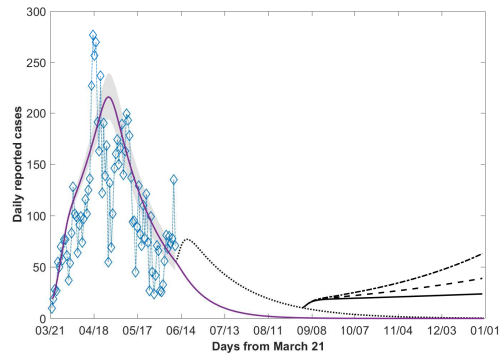}
\caption{Time series plot for model (\ref{eq:Age}), starting from March $21,2020,$
of the numbers of reported individuals change using different lifting
rates at different times as follows: $30\%$ of children and $30\%$
of adults are lifted on June $10$ (dot line) and a fraction $\gamma$
of the remaining confined individuals ($\left(\gamma_{a}=0.115,\gamma_{c}=0.8\right),$
$\left(\gamma_{a}=0.25,\gamma_{c}=0.8\right),$ and $\left(\gamma_{a}=0.39,\gamma_{c}=0.8\right),$
solid line, dashed line, dot-dashed line respectively) is lifted on
September $1st.$}
\label{fig:Age} 
\end{figure}
Although it is believed that the discrete structured-age model (\ref{eq:Age})
is more realistic, our investigations show that this model is the
least accurate of any of the models used to estimate the basic reproduction
number $R_{0}$ since its $SSR$ is the highest one with $SSR=98.76$.
This lead us to think to extend the proposed models by gathering both
age and constant delay factors at once. On the other hand, using our
proposed models, the examination of the lockdown lifting scenario
shows a prominent difference between disease predictions. Furthermore,
no eradication of COVID19 disease is observed before the end of the
year when relaxing the compulsory lockdown on September 1st. The discrete
structure-age model (\ref{eq:Age}), which have estimate the lowest
$R_{0}$ value, predict a less severe disease persistence when comparing
with the other models. Lifting 30\% of the total confined population
on June 10th will lead to a slight second wave of infection followed
by a rapid decrease till the eradication of the disease before the
end of the year.
\begin{figure}[H]
\sidecaption 
\includegraphics[scale=0.28]{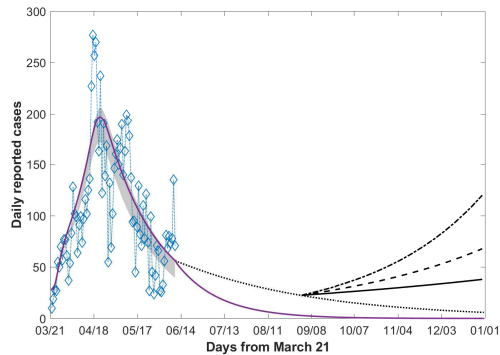}
\caption{Time series plot for model (\ref{eq:delay}), starting from March
$21,2020,$ of the numbers of reported individuals change using different
lifting rates at different times as follows: $30\%$ of the total
population is lifted on June $10$ (dot line) and a fraction $\gamma$
of the remaining confined individuals ($\gamma=0.3,$ $\gamma=0.4,$
and $\gamma=0.5,$ solid line, dashed line, dot-dashed line respectively)
is lifted on September $1st.$}
\label{fig:DDE} 
\end{figure}
However, if this strategy is accompanied with a second lockdown lifting
of at least 30\% of the total confined population (11.5\% of adults
and 80\% of children under 15 years old) on September 1st then the
extinction of the virus cannot happen ( Fig. \ref{fig:Age}) and a
third wave could arise. However, it is obvious that the reopening
of primary and junior high schools does not lead to an important wave
of infection, when comparing with models (\ref{eq:ODE}), (\ref{eq:delay})
and (\ref{eq:Threshold delay}). Figures (\ref{fig:ODE}), (\ref{fig:DDE})
and (\ref{fig:SDE}) show, however, that no second wave will reoccur
but a longer lasting persistence of the infection occurs when 30\%
of the population lifted on June 10th.
However, when comparing with the both basic model (\ref{eq:ODE})
and constant delay model (\ref{eq:delay}), threshold-type delay model
(\ref{eq:Threshold delay}) show a higher size of reported cases as
well as an important second wave when a second stage of lifting lockdown
strategy occurs on September 1st. Furthermore, it seems that the latency
period influence the model fitting to data. Figure (\ref{fig:Effect})
and the $SSR$ measure ($SSR_{\tau=0.33}=92.63$ ), related to the
delay model (\ref{eq:delay}), shows that the constant delay model
with latency period of $8$ hours is the best fit to data and, thus,
this period gives a better prediction than 6 or 12 hours of latency.

In summary, there are no mathematical models able to correctly capture
all complexity of COVID-19 disease in general. Each model, either
simple or complex, has its own advantages and disadvantages. Besides
the availability of data, the choice of model depends on the goal
sought by scientists to answer a question of interest. Furthermore,
the use of complex models does not necessarily provide the most precise
answers than the simplest. Indeed, since many biological and epidemiological
issues related to SARS-CoV-2 remain to be clarified, parameters considered
in the proposed model can be underestimated or overestimated and,
consequently, can lead to wrong results. An unsuccessful evaluation
of the disease behavior could cost serious damage because it lead
to an incorrect estimate of the control health measures that are necessary
to contain the disease transmission. However, although a lot of issues
must be considered to provide the built model a maximum of realism,
mathematical modeling remains a crucial tool to understand and control
the behavior of COVID-19 disease.
\begin{acknowledgement}
The authors are grateful to the anonymous referees for their valuable
and helpful comments that improved this chapter.
\end{acknowledgement}

\end{document}